\documentclass[%
 reprint,
 amsmath,amssymb,
 aps,
]{revtex4-2}
 \usepackage{graphicx}
\usepackage{dcolumn}
\usepackage{bm}
\usepackage{comment}

\begin{document}

\preprint{APS/123-QED}

\title{Engineering monstar polarization disclination through geometric phase}

\author{Ver\'onica Vicu\~na-Hern\'andez}
\author{Pegah Darvehi}%
\altaffiliation[Also at ]{Istituto di Scienze e Tecnologie per l’Energia e la Mobilità Sostenibili, Consiglio Nazionale delle Ricerche, P.le Tecchio 80, 80125 Napoli, Italy.}
\author{Lorenzo Marrucci}%
\altaffiliation[Also at]{CNR-ISASI, Institute of Applied Science and Intelligent Systems, Via Campi Flegrei 34, 80078 Pozzuoli (NA), Italy}
\author{Bruno Piccirillo}%
\email{bruno.piccirillo@unina.it}
\altaffiliation[Also at]{INFN, Sez. di Napoli, Complesso Universitario di Monte Sant'Angelo, via Cinthia, 80126 Napoli, Italy}
\affiliation{%
 Dipartimento di Fisica ``Ettore Pancini'', Universit\`a  degli Studi di Napoli Federico II, Complesso Universitario di Monte Sant'Angelo, Via Cintia, 80126 Napoli, Italy
}%

\date{\today}

\begin{abstract}
We present a method to generate monstar singularities via Pancharatnam-Berry phase by the coherent collinear superposition of two Free-Form Dark Hollow beams, $FFDH^{m}_{q}$, of topological charge $q$, and order of symmetry $m$. FFDH beams are generated with the geometrical parameters of a closed curve exploited to obtain a nonuniform rotation rate of the local polarization azimuth and generate monstar patterns. We report space-variant polarization patterns: radial- and azimuthal-like, lemon- and star-like for symmetric disclinations, and the asymmetric monstar disclination. We present theory and measurements, and find excellent agreement between the two.
\end{abstract}

\maketitle

\section{Introduction}
The generation of classical and quantum states of light is the result of the ability to give structure to the degrees of freedom of light. It has served and impacted in a wide variety of applications, in domains ranging from basic science to manufacturing to communications~\cite{Forbes2021}. Recently, the study of topological features of light has been increased giving rise to discoveries and descriptions about wavefront dislocations~\cite{Berry:01,gbur2015singular,Galvez2014}, and phase and polarization singularities~\cite{Torres2011}. In particular the inhomogeneous space-variant polarization structures in paraxial and non-paraxial regimens that take place in anisotropic media have been investigated in the last years due to their riches in spin-orbit optical interactions~\cite{Cardano2015}. Poincar\'e beams (PBs) are beams of light with inhomogeneous space-variant polarization distribution containing rotational dislocations in directional order~\cite{Galvez:12}. They compass different spatial regions on the Poincar\'e sphere surface over their transverse plane~\cite{Beckley:10}. The singular point of the dislocation is a V- or C-point. Polarization singularities arise when one of the parameters that defines the state of polarization of light is undefined. V-point singularities are related to vectors fields in which the linearly polarized optical field has both orientation and handedness undefined. Poincare-Hopf index $\eta$ defines the order of the V-point singularity~\cite{SP2018}. C-point singularity occurs in ellipse fields and is associated with circular polarization in which the azimuth of the state of polarization is undefined and the orientation of the surrounded ellipses rotate clockwise or counterclockwise about it~\cite{SP2018}. Such rotation is represented by an integer or half-integer index $I_C$, indicating that the azimuthal coordinate on the Poincar\'e sphere rotates $2 |I_C|$ times per turn about the singularity. Basic polarization disclinations are classified as lemons ($I_C=1/2$, 1 radial line), stars ($I_C=-1/2$, 3 radial lines) and monstars ($I_C=1/2$, 3 radial lines). Another characteristic of the disclination is the number of radial lines, L-lines. These are straight lines that originate at the singularity. The polarization orientation is radial at all points along a radial line. Lemons with $I_{C}=1/2$ and stars with $I_{C}=-1/2$ have symmetric patterns with one and three radial lines respectively. Monstar disclination is asymmetric and occurs when the azimuth rotation rate about a C-point exceeds a threshold value and produces more radial lines than lemons or stars with the same $I_C$ index~\cite{Freund:11,Khajavi:16,Cvarch:17}. 
Different approaches have been taken in order to generate the three types of basic disclinations patterns in optical beams. In the majority of cases, the methods adopted are based on the coherent superposition of two distinct spatial modes in two orthogonal polarization states. However, distinct approaches have been exploited depending on the actual tools used for molding wavefronts: sub-wavelength structures~\cite{Niv:05,Beresna:11}, stress birefringence~\cite{Beckley:10}, interferometer-based superpositions~\cite{Maurer:07,Galvez:12,Kumar:13,Vyas:13,Moreno:14,Otte:16}, q-plates~\cite{Cardano:13,Piccirillo:13,Cvarch:17}. In particular the generation of monstar singularities requires the rupture of symmetry of the circular rotation rate of the polarization ellipses semi-major axis. The superpositions of two modes carrying orbital angular momentum (OAM) in orthogonal states of circular polarization breaks off the symmetry by varying the asymmetry of one the optical vortices~\cite{Khajavi_2016}. Other methods of monstar generation have used geometric phase optical elements as elliptically-symmetric q-plates~\cite{Cvarch:17} and Spatially Varying Axis retardation waveplates (SVAPs)~\cite{Piccirillo:20,Darvehi_2021}. Q-plates are nematic liquid-crystal cells where a disclination of order $q$ is patterned by the alignment direction of the liquid crystal molecules called 'director'. The disclination pattern orientation is a function of the angular coordinate $\phi$. Symmetry dislocations as lemons and stars have been generated with q-plates devices with circular symmetry in the orientation of the director pattern. Elliptically-symmetric q-plates and SVAPs have non-circularly symmetry disclination patterns and their director pattern follows a $m$-fold rotational symmetry encoded in the form of the Pancharatnam-Berry phase. The rotational symmetry of order $m$ and other geometrical properties encoded into the Pancharatnam-Berry phase are able to add radial lines to the polarization cross-section structure of the light beam. If $\theta$ is the orientation of polarization at a particular point, and $\phi$ is the angular coordinate, the radial orientation is defined as the orientation of the disclination relative to the radial direction~\cite{Khajavi:16, Galvez:17}:

\begin{equation}
    \theta_{r}= \theta - \phi\,.
\end{equation}

Radial lines occur when $\theta_{r}=n\pi$, with $n$ integer. Radial lines mark the boundaries of angular sectors. The patterns contained within the angular sectors have three types of lines classified as elliptic, hyperbolic, and parabolic. The radial orientation in these sectors change from line to another. Elliptic regions have lines that begin and end at the C-point and the change of radial orientation, $\delta \theta_{r}$, is $+\pi$. Lines in hyperbolic regions avoid the C-point and $\delta \theta_{r} = -\pi$. Parabolic regions appear as a characteristic in monstar patterns. In theses regions one end of the lines connects with the C-point. The number of radial lines for lemons and stars follows the equation~\cite{Khajavi:16, Galvez:17}:

\begin{equation}
\label{eq:numbL}
    N= |2(I_{C}-1)|.
\end{equation}  
Lemons have the rate of rotation of the semi-major axis constant along a circular path centered about the C-point. They have only one direction where the semi-major axis is radial: the semi-major axis rotates at half the rate of circulation about the C-point. Stars rotate in the sense opposite to the path, the axis is radial along three lines. Monstar pattern is more general than lemon and star because in these patterns the semi-major axis has not a constant rate of rotation and the axis may be radial along three or more lines. It contain parabolic sectors delineated by the separatrices, where all the lines of curvature have the C-point as an end point. Monstar patterns have a number of radial lines that do not follow Eq.~\ref{eq:numbL}, and contain parabolic sectors in addition to one or more sectors that can be either parabolic or elliptical. Monstars patterns with positive-, negative- and zero-indexes have been reported generated with non-separable superpositions of three spatial modes carrying orbital angular momentum (OAM) in opposite states of circular polarization~\cite{Galvez:12,Galvez:13,Khajavi_2016}. The criterion of monstar generation has being retaining the index of lemons and stars while adding more radial lines and sectors to invalidate the Eq.~\ref{eq:numbL}.  
In this article we show that the structural features of monstar dislocations can be generated by the Spatially Varying Axis Waveplate (SVAP) device. Monstar singularities require to manipulate the rotation rate of the polarization azimuth around the C- or V -points. SVAP is a device in which the function of the disclination pattern orientation draws the background of a beam carrying OAM with topologial charge $q$ in superposition with a phase modulator factor $\Phi(\phi)$ which is the unitary normal vector orientation of a closed curve~\cite{Piccirillo:20,Darvehi_2021}. This method is able to engineer and set on demand the rotation rate and the structure of the local polarization azimuth around the singularity. The azimuth orientation information is actually encoded in the form of Pancharatnam-Berry phase. SVAP device was introduced in the design and generation of the Free Form Dark Hollow ($FFDH^{m}_{\ell}$) beams of topological charge $\ell$ and symmetry order $m$. It was used to the generating of exotic polarization-structured beams~\cite{Piccirillo:20,Darvehi_2021}.
Non-separable superpositions of FFDH in orthogonal states of polarization produce beams of light that have spatially variable polarization. They map the Poincar\'e sphere onto its transverse mode and carry disclinations by design and have been used to produce high-order disclinations. FFDH can easily produce symmetric lemons and stars. In this work we demostrated the production of monstars with non-separable superposition of FFDH by increasing the number of radial lines (L-lines) and angular sectors in lemon- and star-like patterns while the index $I_{C}=-1/2$ is maintained with the same value. 
We illustrate the operation of SVAPs encoding geometric phase distribution with a non-uniform azimuthal dependence, explain how we use them to create various polarization patterns and how to modulate its parameters in order to get monstar singularities.  we introduce some examples of such beams and demonstrate their properties on experimental grounds.

\subsection{Operation of SVAPs}
\label{subs:operationSVAP}

The operation principle of SVAP devices has been already described elsewhere~\cite{Piccirillo:10,Piccirillo:13,AlemanCastaneda:19,Rubano:19}. In particular, the optical effect of a generic Free-Form Azimuthal SVAP can be described by the operator $\hat{Q}(q, m,\xi)$ acting on the local polarization state of the field, defined as
\begin{eqnarray}\label{eq:FFSVAP}
\hat{Q}(q,m,\xi)&=&\cos{\frac{\delta}{2}}\left(|L\rangle \, \langle L | + |R\rangle \, \langle R |\right) \nonumber\\
&&+ {\rm i}\sin{\frac{\delta}{2}}\left({\rm e}^{{\rm i} 2 \Psi(\phi;q, m,\xi)}|R\rangle  \langle L| \right) \nonumber\\ 
&&+  {\rm i}\sin{\frac{\delta}{2}} \left({\rm e}^{-{\rm i} 2 \Psi(\phi;q, m,\xi)} |L\rangle  \langle R | \right),
\end{eqnarray}
in which the phase factor ${\rm e}^{{\rm i}\, 2\, \Psi(\phi;q, m,\xi)}$ corresponds to Pancharatnam-Berry phase function of the parameters: topological charge $2q$, phase modulation $\phi$, $m$-fold symmetry of a closed curve $\rho$, and $\xi$ the geometrical properties of the closed curve $\rho$. All these parameters are directly mapped to the generated Free-Form Dark Hollow, $FFDH^{m}_{\ell}$ beam. Our device SVAP allows the electrical control of the phase retardation (birefringence) $\delta$, which can be set at $\delta=\pi$ called half wave plate configuration HWP or at $\delta=\pi/2$ the quarter wave plate QWP configuration. 
Let us consider the two operation modes of SVAP device and linear and circular input polarization to obtain four polarization structures: radial-like, azimuthal-like, lemon-like, and star-like. In the first case, the SVAP retardation is set to $\delta=\pi/2$ (quarter-wave plate operation) and the input light is a circularly polarized Gaussian beam, the output beam is a coherent superposition of a Gaussian beam and a beam carrying the geometric phase with orthogonal polarization respect to the input beam. These beams are so-called Modulate Poincar\'e Beams due to their diversity in polarization singularity landscapes~\cite{Darvehi_2021}. In the near field, the resulting MPB has a linear polarization map just point-by-point coincident with the SVAP optic axis distribution, with a central C-point of index $I_C=q$. A second structure of the polarization map can then be obtained by using a haf-wave plate after the SVAP to exchange the handedness of the circular components of the MPB. SVAP device reverses the circular polarization direction, introduce Orbital Angular Momentum (OAM) by $2q\,\phi$, and the modulation phase factor giving by $q\,\Phi(\phi)$. In the third configuration mode, the SVAP retardation is set to $\delta=\pi$ (half-wave plate operation) and the input polarization is linear. SVAP transforms linearly polarized light into a coherent superposition of right- and left-handed polarized light with opposite OAM and an additional $m$-fold rotation-symmetric phase values, which add at every point in the beam transverse plane leading to a vector field with a central V-point with $I_V=2q$. The MPB vector field is of course also dependent on the input linear polarization orientation, which introduces a further local rotation of the polarization by a uniform angle giving as output either radial- or azimuthal-like polarizations. 

\begin{figure}[ht!]
\begin{center}
\includegraphics[width=8cm]{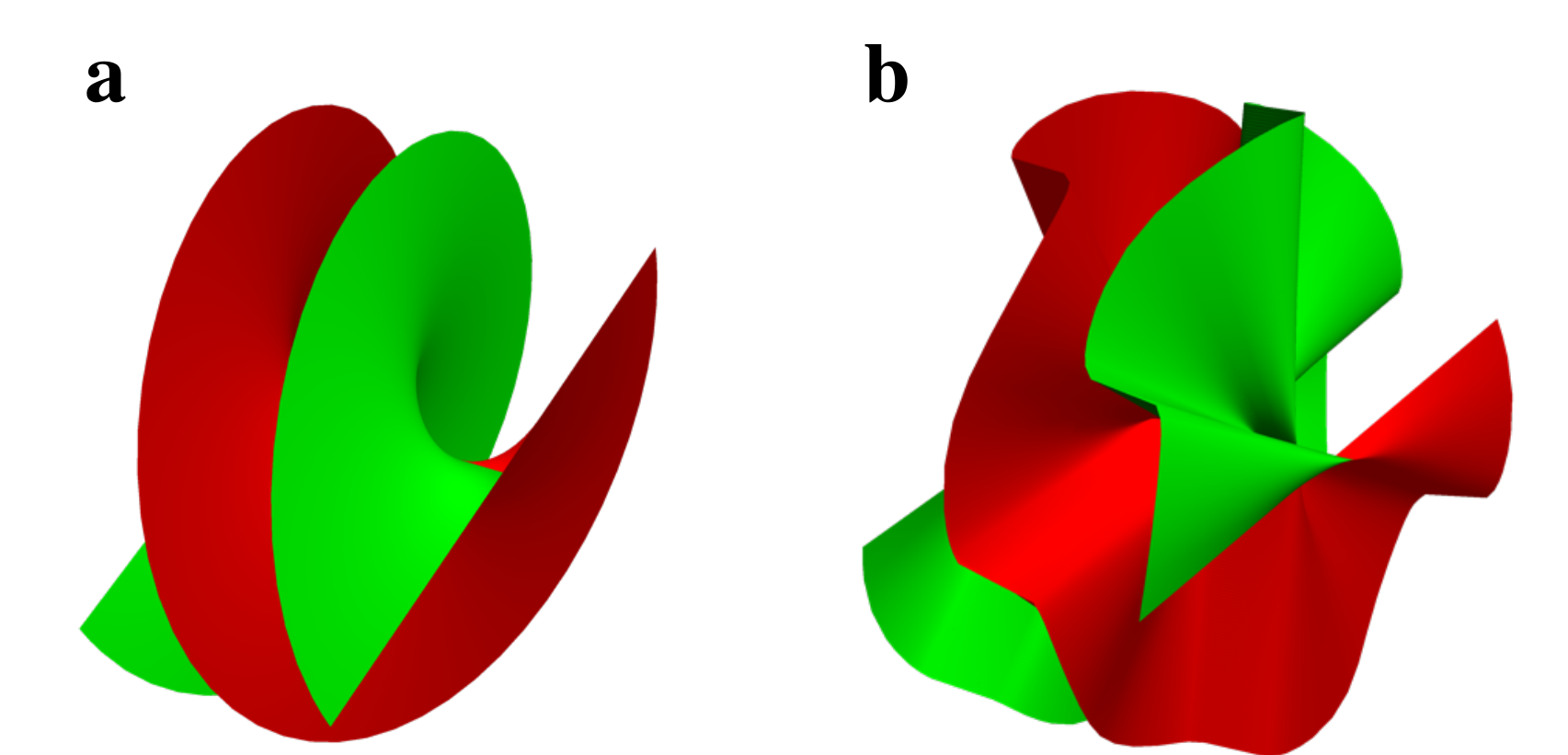}
\end{center}
\caption{Modulated vs unmodulated helical wavefronts: ({\bf a}) uniform helical wavefront for $\ell=2$ ($\gamma_{\infty}$); ({\bf b}) modulated helical wavefront for $\ell=2$ and $\gamma_5(a=b,n_1=1/2;n_2=n_3=4/3)$.}\label{fig:Fig1}
\end{figure}

An FFH spatial mode of $m$-order and topological charge $\ell=2q$ is generated by adding an additional $m$-fold rotation-symmetric phase, $\bar{\psi}(\phi)$ to a beam carrying OAM with topological charge $\ell$ as we can see the pictured example in Fig.~\ref{fig:Fig1}. This additional phase gives a nonuniform distribution of the phase with an  $m$-fold symmetry structure. The modulation azimuthal phase factor, $e^{{\rm i}\, q\,\bar{\psi}(\phi)}$  ~\cite{Piccirillo:20}, corresponds to the optic axis distribution, which is giving by the orientation of the normal vector $\mathbf{\hat{n}}$ along the closed curve $\rho (\phi)$.\\
The Pancharatnam–Berry phase for the SVAP device is:
\begin{equation}\label{eq:FFH_phase}
 e^{{\rm i}2\Psi(\phi;q, m,\xi)}=e^{{\rm i} \ell \vartheta(\phi)},   
\end{equation}
where $\vartheta$ is defined as the polar angle of the unit vector $\hat{\boldsymbol{n}}=(n_x,n_y)$ which reads:
\begin{eqnarray}\label{eq:nxny}
n_x(\phi)=\cos\vartheta=\frac{\rho\cos\phi+\dot{\rho}\sin{\phi}}{\left(\dot{\rho}^2+\rho^2\right)^{1/2}}, 
\nonumber\\
n_y(\phi)=\sin\vartheta=\frac{\rho\sin\phi-\dot{\rho}\cos{\phi}}{\left(\dot{\rho}^2+\rho^2\right)^{1/2}},
\end{eqnarray}
the closed curve and its first derivative read:
\begin{eqnarray}\label{eq:rho}
\rho(\phi)&=&
\left( \left| \frac{\cos{\frac{m \phi}{4}}} {a} \right |^{n_2} + \left| \frac{\sin{\frac{m \phi}{4}}}{b}  \right |^{n_3} \right)^{-\frac{1}{n_1}},
\nonumber\\
\dot{\rho}&=&\frac{d\rho}{d\phi}.
\end{eqnarray}
The Eq.~(\ref{eq:rho}) represents the superformula in polar coordinates for multiple classes of plane curves $\gamma_m(a,b,n_1,n_2,n_3)$ of the most diverse kinds~\cite{Gielis:03}. Specifically $\rho$ is the distance of a point of the curve $\gamma_m$ from the origin of the coordinate system as a function of the azimuthal angle $\phi$, $m$ is an integer number, $n_1$, $n_2$ and $n_3$ are three integers controlling its local radius of curvature and, finally, the positive real numbers $a$ and $b$ parameterize the radii of the circumferences respectively inscribed and circumscribed to the curve $\gamma_m$. It is worth noticing that, for $m=4$, $a=b$ and $n_2=n_3>2$, the superformula simply returns the superellipses first introduced by G. Lam\'e in 1818~\cite{Hazewinkel01}.\\
The phase of an FFH mode is hence proportional to the angle $\vartheta(\phi)$ formed by the unit vector $\hat{\boldsymbol{n}}$ normal to the curve $\gamma_m(a,b,n_1,n_2,n_3)$ with respect to a specified reference axis $x$.\\ 

An $\ell$-charge $m$-order FFH mode obviously does not carry a well-defined OAM and its spectrum includes multiple indices reflecting the $m$-fold rotational symmetry of the generating curve. In order to determine such indices, it is convenient to express the phase $\Psi(\phi)$ as follows
\begin{equation}\label{eq:FFH_phase2norm}
e^{{\rm i}2\Psi(\phi)}=\left(n_x + {\rm i} \, n_y\right)^{\ell}=\left[\frac{\rho(\phi) - {\rm i}\, \dot{\rho}(\phi)}{\rho(\phi) + {\rm i}\, \dot{\rho}(\phi)}\right]^q e^{{\rm i}\,2 q\,\phi}=e^{{\rm i}\, q\,\bar{\psi}(\phi)}\,e^{{\rm i}\,2 q\,\phi},
\end{equation}
where we set $q=\ell/2$. $q$ is hence integer or half-integer depending on whether $\ell$ is even or odd, respectively. The FFH phase factor therefore can be separated into the product of the helical background phase factor $e^{{\rm i}\ell\phi}=e^{{\rm i}\,2 q\,\phi}$ times the modulation phase factor $e^{{\rm i}\, q\,\bar{\psi}(\phi)}$. The latter is a periodic function of $\phi$ with period $2\pi/m$, i.e. 
\begin{equation}\label{eq:FFH_periodical}
e^{{\rm i}\, q\,\bar{\psi}(\phi+\frac{2\pi}{m})}=e^{{\rm i}\, q\,\bar{\psi}(\phi)},
\end{equation}
and hence can be expanded in a Fourier series
\begin{eqnarray}\label{eq:FFH_Fourier}
e^{{\rm i}\,q\,\bar{\psi}(\phi)}&=&\sum_h {\chi_h} e^{{\rm i}\, h\, m\,\phi},\nonumber\\
{\chi_h}&=&\frac{m}{2\pi}\int_0^{\frac{2\pi}{m}}{\left[\frac{\rho(\phi) - {\rm i}\, \dot{\rho}(\phi)}{\rho(\phi) + {\rm i}\, \dot{\rho}(\phi)}\right]^q\, e^{-{\rm i}\, h\, m\,\phi}\,d\phi}.
\end{eqnarray}
Then, the OAM spectrum of an $\ell$-charge $m$-order FFH beam turns out to include only the components with indices $\ell \pm h\, m$, $h$ being any integer and $\ell=2q$ the OAM index corresponding to the background helical mode. The azimuthal phase factor in Eq.~(\ref{eq:FFH_phase_exp}) can be finally expressed as the following helical mode expansion
\begin{equation}\label{eq:FFH_phase_exp}
e^{{\rm i}2\Psi(\phi)}=\sum_h \chi_h e^{{\rm i}\, (2 q + h\, m\,\phi)}=\sum_l c_l e^{{\rm i}\, l \,\phi},\hspace{1cm}l=2 q + h\, m.
\end{equation}
The mean value of the OAM in an $m$-order FFH mode can be easily calculated from Eqs.~(\ref{eq:FFH_phase}) and (\ref{eq:nxny}),
\begin{eqnarray}\label{eq:OAM}
\langle L_z \rangle &=&-\frac{{\rm i}\hbar}{2\pi}\int_{0}^{2\pi}{e^{-{\rm i}2\Psi(\phi)}\,\frac{\partial}{\partial \phi} \,e^{{\rm i}2\Psi(\phi)}} \nonumber\\
&=&\frac{\hbar}{\pi}\,\left.\Psi(\phi)\right|_0^{2\pi}
=\frac{\hbar}{\pi}q \left.\vartheta(\phi)\right|_0^{2\pi} = 2\hbar q.
\end{eqnarray}
This result indicates that the mean OAM carried by an $\ell$-charge FFH mode is always $\hbar \ell = 2 \hbar q$ per photon, independently of $m$. It is therefore coincident with the OAM per photon of the background helical mode and it is proportional to the overall topological charge of the transverse wavefront. The phase modulation factor $e^{{\rm i}\,q\,\bar{\psi}(\phi)}$ defined in Eq.~(\ref{eq:FFH_phase2norm}) has a zero mean OAM. However, the specific value of the fraction $|c_l|^2$ of the total power of the optical field carrying an OAM proportional to $l$ depends on the geometric details of the generating curve $\gamma_m(a,b,n_1,n_2,n_3)$ and ultimately on the specific values of the parameters in Eq.~(\ref{eq:rho}). Uniform helical modes can be easily obtained as a special case of FFH modes when the generating curve $\gamma_m$ degenerates into a circumference ($\gamma_m \rightarrow \gamma_{\infty} $).
Free space propagation obviously alters the shape of FFH modes, but it does not influence their symmetry properties.

\begin{figure}[htbp]
\centering
\includegraphics[width=\linewidth]{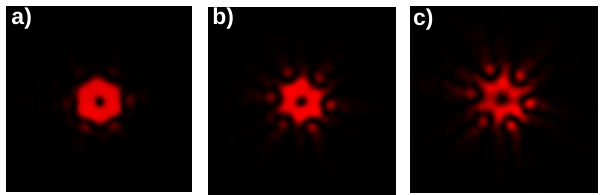}
\caption{Transverse intensity of the beams generated by the SVAP devices designed with a 6-fold symmetry, $n_{2}=n_{3}=2.3$, $a=b$ and different number of L-lines modulated by the parameter $n_{1}$. a) Profile 1 has $n_{1}=2/3$. b) Profile 2 has $n_{1}=3/10$. c) Profile 3 has $n_{1}=1/6$.}
\label{fig:TI_profiles}
\end{figure}

In this work we introduce three examples of SVAP device with parameters $m=6$, $n_2=2.3$, $n_3=2.3$, $a=b$ and distinct parameters $n_{1}$ for each distinct SVAP profile. With these examples, we illustrate the ability of the SVAP device to produce monstar singularities by adding radial lines and angular sectors to the polarization structure that originally performs as a lemon-like or as a star-like . The three profiles samples maintained the same $m$-fold rotational symmetry and also the parameters $a=b$, $n_{2}=n_{3}$, only the parameter $n_{1}$ was set distinct for each profile. Parameter $n_{1}$ is able to modulate the radius of curvature of $\gamma_m(a,b,n_1,n_2,n_3)$, while  $1/n_{1}$ increases in value, then the radius of curvature of $\gamma_m(a,b,n_1,n_2,n_3)$ decreases respect to the origin of the coordinate system. It seems as if the beam were compressed towards the center. In Fig.~\ref{fig:TI_profiles} can be observed the transverse intensity of the beams generated with different parameter $1/n_{1}$. The beam is more compressed as the value of $1/n_{1}$ increases and also, the number of radial lines L-lines increases as well. We show this in the section of results and discussion.

\section{Experiment}


\begin{figure}[ht!]
\centering\includegraphics[width=\linewidth]{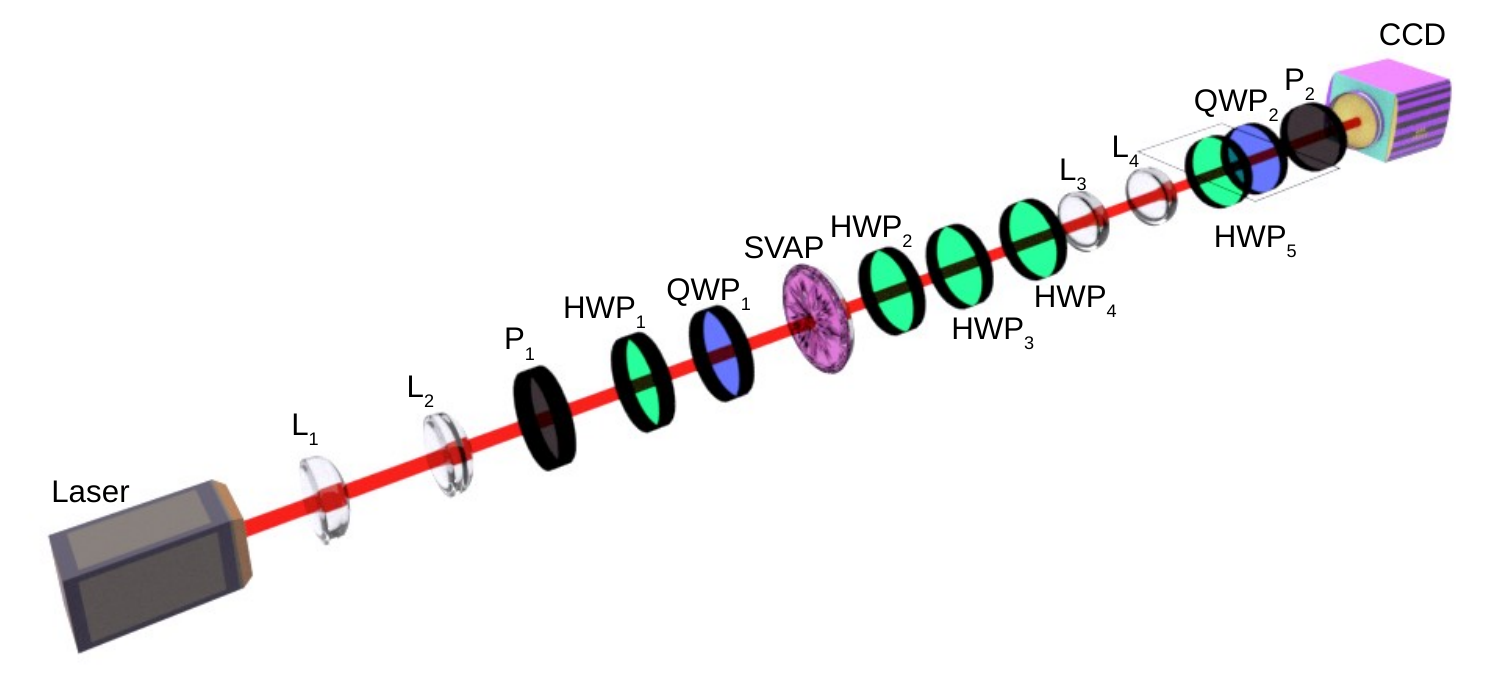}
\caption{Experimental setup for the measurements in near field.}
\label{fig:exp_setup_near_field}
\end{figure}

Our liquid crystals SVAPs have been fabricated adopting a ``direct-write approach'' to pattern the LC anchoring to the ITO-coated glass walls of the cell. Electrical control of the retardation $\delta$ -- or electrical tunability -- and non-diffractive operation are just some of the main advantages of this technology. The pattern of each profile was generated by following the orientation respect to the $x$-axis of the normal vector of the 6-fold symmetry curve as it was described in section~\ref{subs:operationSVAP}. 

To generate the beams with spatially varying polarization we use a beam from a Helium-Neon laser centred at $\lambda=633$nm with a beam diameter of ~0.2mm. The Fig.~\ref{fig:exp_setup_near_field} shows the experimental setup. The beam is sent through the lenses L1 and L2 with focal lengths of $f_{1}=$5 cm and $f_{2}=$20 cm and separated by 25 cm forming a telescope T$_{1}$ to magnifying the beam by a factor of ~4$\times$. A linear polarizer (P$_{1}$) was placed to orientate the beam into linear polarization. We used a quarter wave plate (QWP$_{1}$) or a half wave plate (HWP$_{1}$) to obtain either circular or linear input polarization. The polarized beam impinges on the SVAP device. SVAP turns the Gaussian beam into a structured beam carrying the Pancharatnam-Berry phase. To generate the spatially-variable polarization singularities, the retardation $\delta$ must be put in half wave configuration $\delta = \pi$ (HWP) or quarter wave plate configuration $\delta = \pi/2$ (QWP) to get  the coherent superposition of two modes with perpendicular polarization.
Our SVAP devices produce spatially varying polarization singular patterns shaped by the symmetry properties of the curve encoded in the Pancharatnam-Berry phase. We measured the transverse polarization distribution in the near field. The light at the plane of SVAP (near field) was re-imaged onto a second plane by two lenses, $L_{3}$ and $L_{4}$, with focal length $f=15$cm in 4-f configuration. The plane of near field was located at the focal distance of L$_{4}$, in which a CCD camera was placed. In order to obtain the state of polarization of the beams, we measured the Stokes parameters with the half wave plate (HWP$_{4}$) and with the QWP$_{2}$ which were placed before of CCD plane and the analyser (P$_{2}$). 

\section{Results and Discussion}

\begin{figure}[htbp]
\centering
\includegraphics[width=\linewidth]{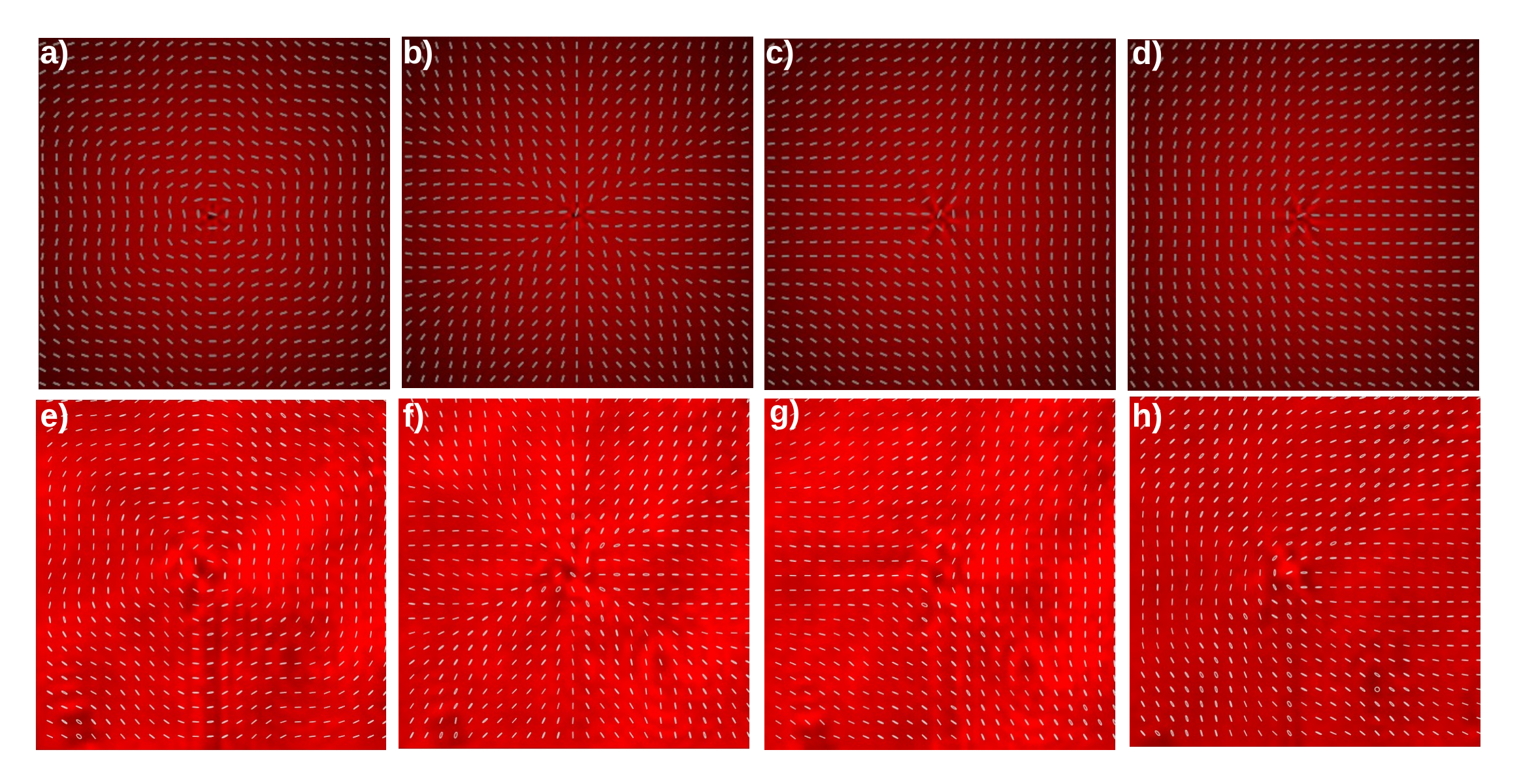}
\caption{Theoretical and experimental measurements of the polarisation structure in near field for the profile 1. First row shows the theoretical results: a) azimuthal-like pattern, b) radial-like pattern, c) star-like pattern, and d) lemon-like pattern. Second row shows the experimental measurements: e) azimuthal-like pattern, f) radial-like pattern, g) star-like pattern, and h) lemon-like pattern.}
\label{fig:nearfield:p1}
\end{figure}

\begin{figure}[htbp]
\centering
\includegraphics[width=\linewidth]{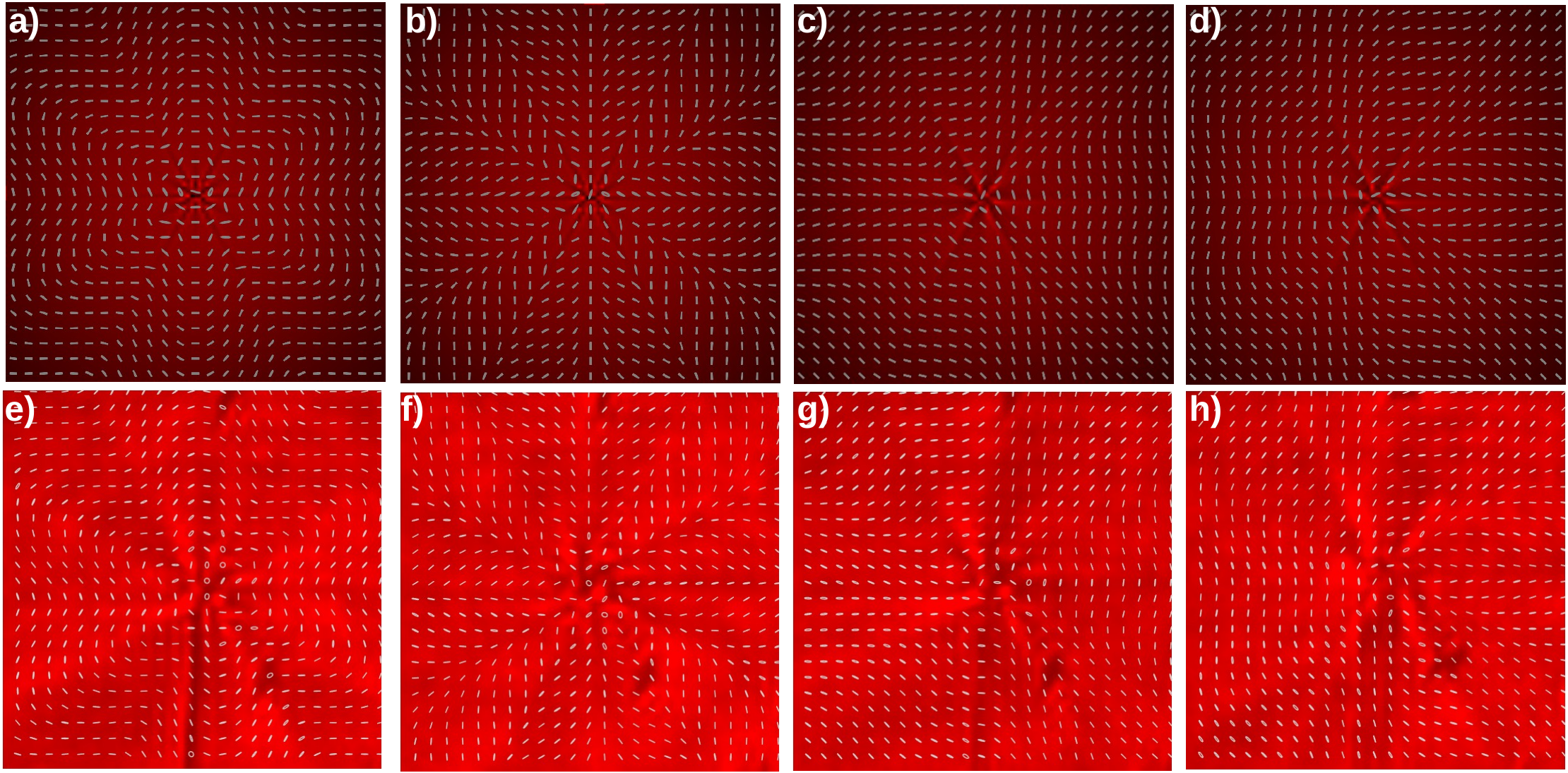}
\caption{Theoretical and experimental measurements of the polarisation structure in near field for the profile 2. First row shows the theoretical results: a) azimuthal-like pattern, b) radial-like pattern, c) star-like pattern, and d) lemon-like pattern. Second row shows the experimental measurements: e) azimuthal-like pattern, f) radial-like pattern, g) star-like pattern, and h) lemon-like pattern.}
\label{fig:nearfield:p2}
\end{figure}

\begin{figure}[htbp]
\centering
\includegraphics[width=\linewidth]{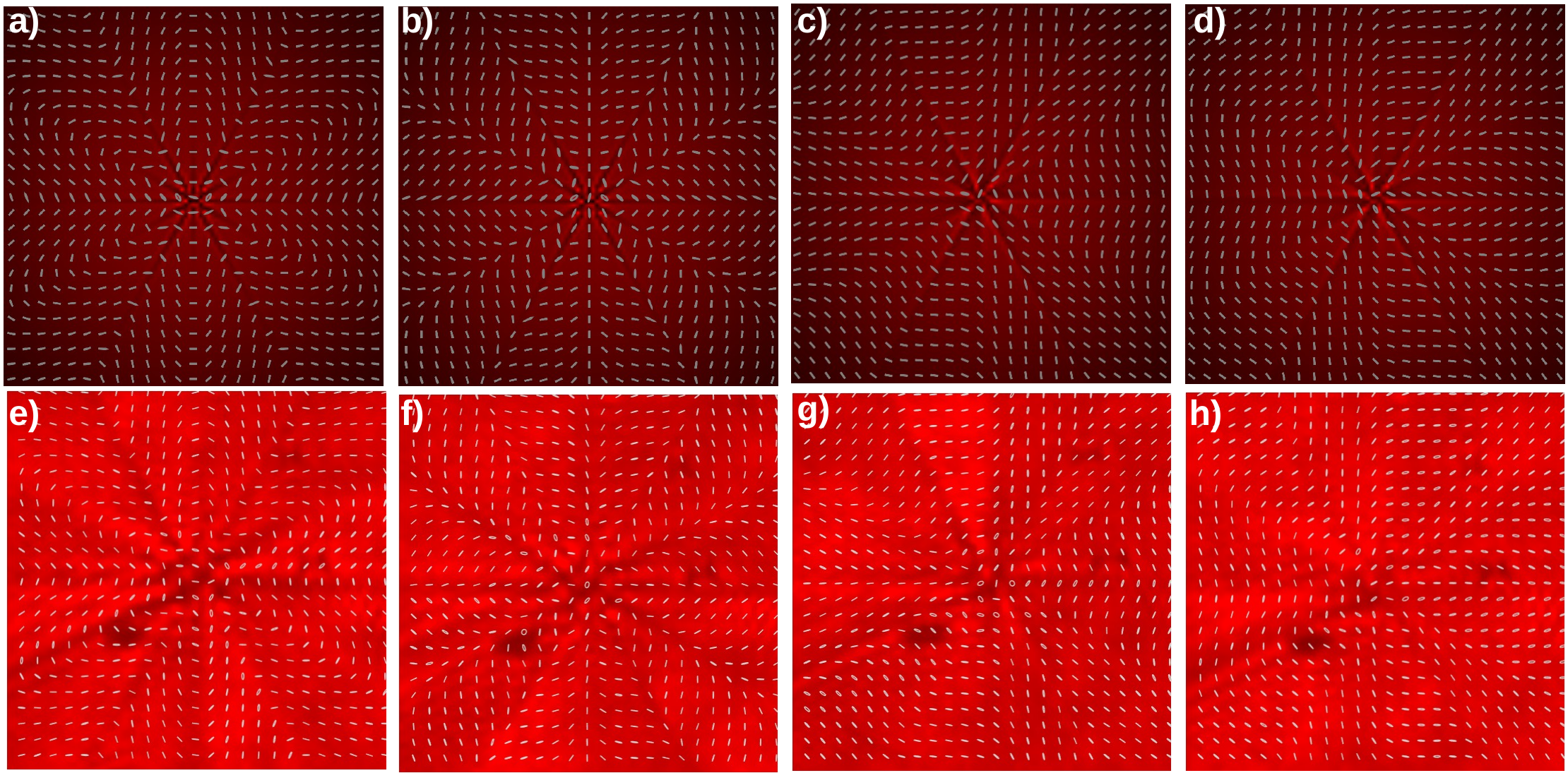}
\caption{Theoretical and experimental measurements of the polarisation structure in near field for the profile 3. First row shows the theoretical results: a) azimuthal-like pattern, b) radial-like pattern, c) star-like pattern, and d) lemon-like pattern. Second row shows the experimental measurements: e) azimuthal-like pattern, f) radial-like pattern, g) star-like pattern, and h) lemon-like pattern.}
\label{fig:nearfield:p3}
\end{figure}

In this section we present the transverse structures of polarization that were obtained by using the three designed SVAP devices. We have called these patterns as radial-like, azimuthal-like, lemon-like, and star-like due to they resemble the basic singularities patterns with the additional m-symmetry phase factor which draws the shape of symmetry on the pattern. We show how to generate monstar singularities by breaking the circular symmetry of the rotation rate of the axis of orientation of the polarization of lemon- and the star-like patterns so that we obtain monstar with positive and negative index $I_{C}$. The intensity distributions showed in Fig.~\ref{fig:TI_profiles} corresponds to the far field, HWP operation mode $\delta=\pi$, and circular input polarization. The parameter $n_{1}$ is associated to the number of radial lines L-lines and angular sectors present in the polarization structure. It appears as the fractional power $\frac{1}{n_{1}}$ in the expression for $\rho(\phi)$ shown in Eq.~\ref{eq:rho} having numerical values for our designed SVAPs; $\frac{3}{2}$, $\frac{10}{3}$, and 6, for profile 1, profile 2, and profile 3 respectively. We can see that by increasing the parameter $\frac{1}{n_{1}}$ the structures of the transverse intensity get narrower, in particular, the radius of curvature measured with respect to the phase singularity is reduced as if the beam were compressed towards the center shaping narrower edges. We obtained the polarization structure in near field for each profile and showed the evolution of the basic-like structures turning into monstar singularities when L-lines and angular sectors are added by increasing the parameter $\frac{1}{n_{1}}$ while the value of index $I_{C}=|1/2|$ is maintained.\\
We show theory and measurements for the four configurations of input polarization and retardance value $\delta$. Azimuthal- and radial-like patterns are performed with retardance $\delta=\,\pi$ and horizontal and vertical input polarization respectively. Lemon- and star-like configurations are performed with retardance $\delta=\pi/2$ and right circular input polarization to obtain star-like pattern. A half wave plate is placed after the SVAP, showed as HWP$_{3}$ in Fig.~\ref{fig:exp_setup_near_field} to flip the handedness of circular polarization and obtain lemon-like pattern. Figures~\ref{fig:nearfield:p1}, \ref{fig:nearfield:p2} and \ref{fig:nearfield:p3} show polarization structure in near field obtained with SVAP of: $\frac{1}{n_{1}}=\frac{3}{2}$ (profile 1), $\frac{1}{n_{1}}=\frac{10}{3}$ (profile 2), and $\frac{1}{n_{1}}=6$ (profile 3) respectively, the first row shows the theoretical patterns and the second row shows the experimental measurements. The patterns appear in the following order: azimuthal-like, radial-like, star-like, and lemon-like. We can observe the evolution of the angle of polarization orientation in the polarizations structures for the three samples with different value of parameter $1/n_{1}$. Modulation phase factor $\bar{\psi}(\phi)$ gives structure to the transverse intensity and polarization. In the azimuthal-like configurations in figures~\ref{fig:nearfield:p1}a), \ref{fig:nearfield:p2}a), \ref{fig:nearfield:p3}a) we observe a structure like flower petals inherited from the 6 order rotational symmetry of the curve $\rho$, and that is stronger as parameter $\frac{1}{n_{1}}$ increases. In particular, the star-like configurations show more radial lines and angular sectors as the parameter $1/n_{1}$ increases turning the usually basic star structure of profile 1 into a more complex structures for profiles 2 and 3 which have six and nine radial lines and angular sectors respectively. These last two are monstar patterns. 

\begin{figure}[htbp]
\centering
\includegraphics[width=\linewidth]{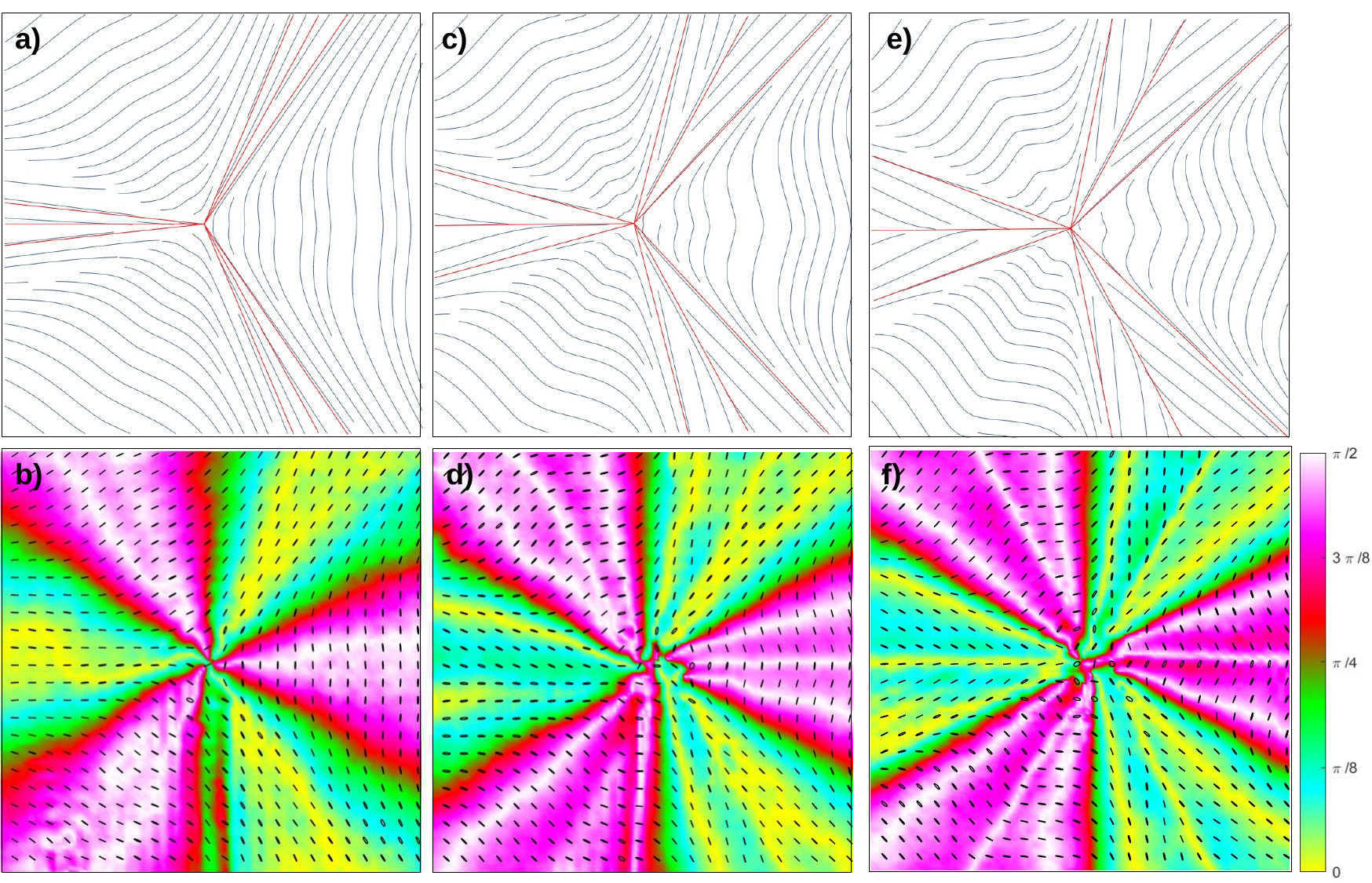}
\caption{Numerical simulations and experimental measurements of the star patterns. False color encodes the orientation of the polarization relative to the radial direction: $\theta_{r}= \theta-\phi$. L-lines are along the yellow colour.}
\label{fig:radial_lines_star}
\end{figure}

\begin{figure}[htbp]
\centering
\includegraphics[width=\linewidth]{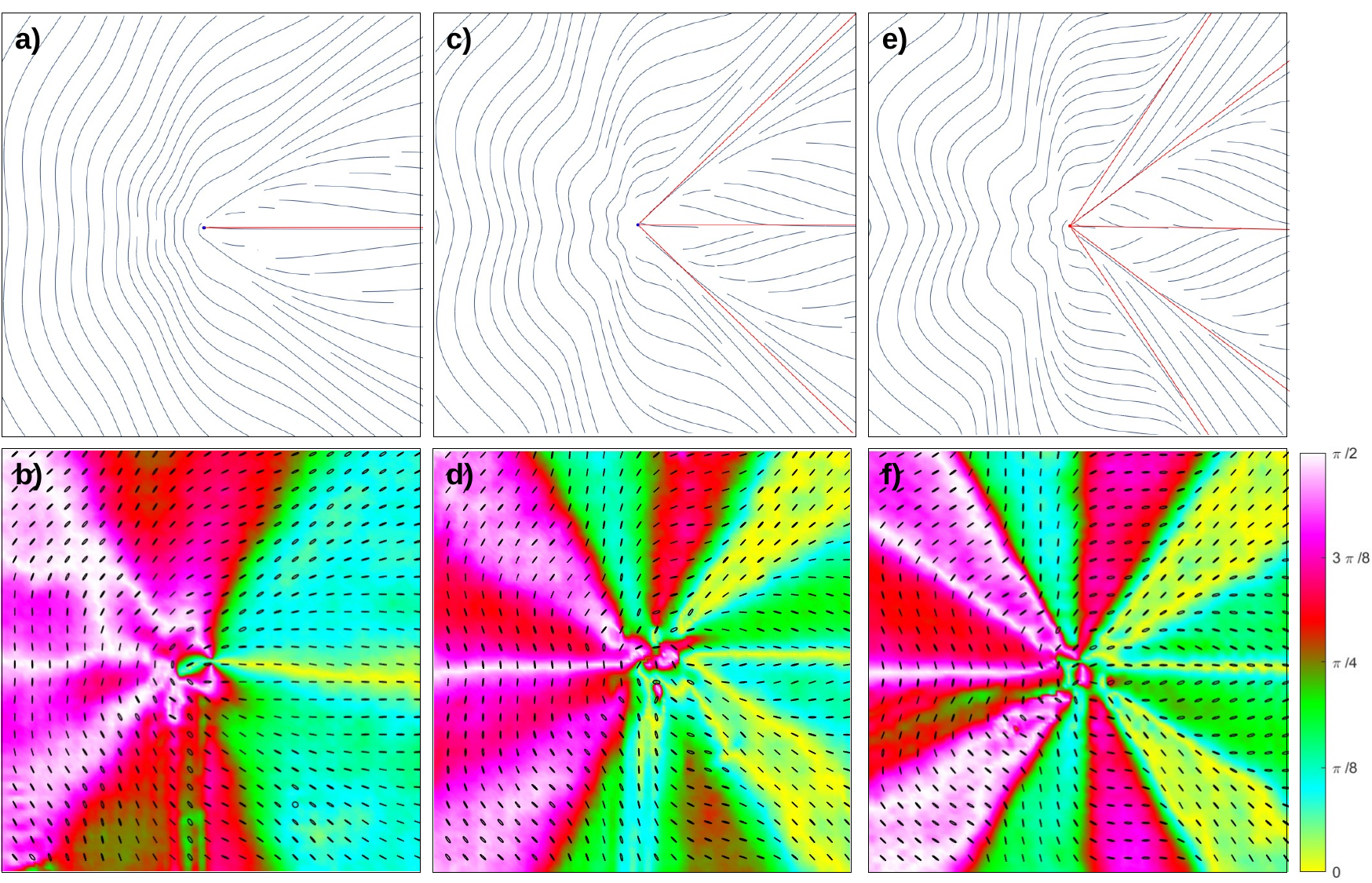}
\caption{Numerical simulations and experimental measurements of the star patterns. False color encodes the orientation of the polarization relative to the radial direction: $\theta_{r}= \theta-\phi$. L-lines are along the yellow colour}
\label{fig:radial_lines_lemon}
\end{figure}

Radial lines are indicated with a yellow background colour in figures~\ref{fig:radial_lines_star} and \ref{fig:radial_lines_lemon}. It shows the patterns star-like and lemon-like of the profiles 1, 2, and 3 with a background colour corresponding to the orientation of polarization relative to the radial direction $\theta_{r} = \theta - \phi$. Radial lines are indicated in yellow colour and angles are in radians. Monstars distinguishes from lemons and stars because their parabolic sectors, delimited by radial lines where the lines radiate from the singularity. A disclination pattern in which the number of radial lines does not follow the equation~\ref{eq:numbL} is defined as a monstar disclination~\cite{Galvez2014}. We can see the transition from the star- and lemon-like patterns to monstar patterns by adding sectors and radial lines by the modulation of the parameter $\frac{1}{n_{1}}$ in each sample of SVAP device. In the star-like Figure~\ref{fig:radial_lines_star}b) is shown the star-like pattern of profile 1 which has $\frac{1}{n_{1}}=\frac{3}{2}$, all sectors are hyperbolic and it has three radial lines as the classical star pattern. In Fig.~\ref{fig:radial_lines_star}d) is shown the star-like pattern of profile 2 which has $\frac{1}{n_{1}}=\frac{10}{3}$, there are six radial lines, it seems as if each radial line of the basic star pattern were split into two. Within these radial lines, there are parabolic sectors that change their radial orientation from one line to another by $\delta_{r} \leq \pi/8$. Finally, Fig.~\ref{fig:radial_lines_star}f) shows the star-like pattern of profile 3 which has $\frac{1}{n_{1}}=6$, in this pattern the parabolic sectors and radial lines split into three and the radial orientation change form one to another by $\delta_{r} \leq \pi/8$. The structure has nine lines and parabolic sectors from one line to another. Figures \ref{fig:radial_lines_star}d) and \ref{fig:radial_lines_star}f) are monstar because the number of radial lines does not follow the equation~\ref{eq:numbL}. These monstars patterns have an index $I_{C}=-1/2$. 

The generation of monstar patterns with the SVAP device allows adding sectors and radial lines by the modulation of the Pancharatnam-Berry phase while maintaining the value of orbital angular momentum (OAM) which is directly related to the index $I_{C}$. 
The transition from lemon patterns to monstar patterns is illustrated in Fig.~\ref{fig:radial_lines_lemon}. Radial lines pointed out with the yellow colour background indicating that relative angle of the polarization orientation and the radial direction is zero. False colour encodes the orientation of the polarization relative to the radial direction $\theta_{r}$. A remarkable characteristic about the monstars in Figures~\ref{fig:radial_lines_star} and \ref{fig:radial_lines_lemon} is that all the lines of curvature in the sectors
delineated by the separatrices, have the C-point as an end point. In lemon-like patterns we can observe how the radial lines and parabolic sectors increase in the patterns of profiles 2 and 3. Profile 2 has for example, two parabolic regions localized at the angular positions: $36^{\circ}$ to $54^{\circ}$, and $-26^{\circ}$ to $-49^{\circ}$, and the usual radial line for lemons localized at zero rads. Profile 3 also has two paraboloc regions localized at the angular positions: $36^{\circ}$ to $49^{\circ}$, and $-36^{\circ}$ to $-57^{\circ}$ and the usual radial line for lemons localized at zero rads. Patterns shown in Figures~\ref{fig:radial_lines_lemon}d) and \ref{fig:radial_lines_lemon}f) are monstar because the number of radial lines does not follow the equation~\ref{eq:numbL}. These monstars patterns have an index $I_{C}=1/2$. The images have a resolution of 60 $\times$ 60 pixels in size. 


\section{Conclusions}

We presented the generation of beams with spatially varying polarization emphasizing monstars patterns through the non- separable coherent superpositions of FFDH beams in orthogonal states of polarization produced by Spatially Varying Axis waveplates SAVP devices that were designed with parameters of geometric properties of a closed curve. We showed how the star- and lemon-like patterns evolve from the basic structures and number of radial lines and angular sectors to more complex structures, and radial lines, and angular sectors which are monstar patterns. It is remarkable that these monstars patterns were generated without changing the index $I_{C}$ to add angular sectors to the basic structures and the rupture of symmetry was done on the rate rotation of the optic axis of the devices which means also a non-symmetric Panchratnam-Berry phase. The rupture of symmetry can also be seen in the rotation rate of the local polarization azimuth around the singularity. We showed theoretical and experimental measurements with excellent agreement.  




\begin{acknowledgments} This work was supported by the European Union (EU) within Horizon 2020—European Research Council Advanced Grant No. 694683 (PHOSPhOR) and by INFN -- Naples, ETHIOPIA (CSN 5).
\end{acknowledgments}

\appendix

\bibliography{references1}

\end{document}